\documentclass[12pt]{article}

\newcommand{\ft}[2]{{\textstyle{\frac{#1}{#2}}}}

\newcommand{\dr}{\raise.3ex\hbox{$\stackrel{\leftarrow}{\partial }$}{}}
\newcommand{\delr}{\raise.3ex\hbox{$\stackrel{\leftarrow}{\delta }$}{}}

\begin{document}
\renewcommand{\theequation}{\thesection.\arabic{equation}}
\csname @addtoreset\endcsname{equation}{section}
\newcommand{\dkt}{\delta_\kappa \theta}
\newcommand{\alphamsbar}{{\alpha_{\overline{\text{MS}}}}}
\newcommand{\be}{\begin{equation}}
\newcommand{\ee}{\end{equation}}
\newcommand{\bea}{\begin{eqnarray}}
\newcommand{\eea}{\end{eqnarray}}
\newcommand{\ba}{\left(\begin{array}}
\newcommand{\ea}{\end{array}\right)}
\newcommand{\eps}{\epsilon_{\mu \nu \rho \sigma} k_1^\mu k_2^\nu k_3^\rho
k_4^\sigma}
\newcommand{\ul}{\underline}
\newcommand{\Dsl}[1]{\slash\hskip -0.20 cm #1}
\newcommand{\al}{\alpha}
\newcommand{\s}{\sigma}
\newcommand{\w}{\omega}
\newcommand{\R}{\rho}
\newcommand{\La}{\Lambda}
\newcommand{\ens}{e^{-n\sigma}}
\newcommand{\eens}{e^{-n\sigma_0}}
\newcommand{\ch}{{\tt (*)}\ }
\begin{titlepage}
\begin{flushright}
KUL-TF-98/23\\ SU-ITP-98/27\\ {\tt hep-th/9804177}\\
April 28, 1998; Revised September 1998.
\end{flushright}
\vfill
\begin{center}
{\Large \bf Black Holes\\[.5cm] and \\[.5cm]
Superconformal Mechanics}
\vskip 0.3cm
{\large {\sl }}
\vskip 10.mm
{\bf Piet Claus$^a$, ~Martijn Derix$^a$, ~Renata
Kallosh$^{b}$, ~Jason Kumar$^{b}$,\\[2mm]
 Paul K. Townsend$^{c,\star}$
{}~and {}~Antoine Van Proeyen$^{a,\dagger}$ } \\
\vskip 0.8 cm
 $^a$
Instituut voor theoretische fysica, \\ Katholieke
Universiteit Leuven, B-3001 Leuven, Belgium\\[2mm]
$^b$
Physics Department, \\ Stanford
University, Stanford, CA 94305-4060, USA\\[2mm]
$^c$ Institute for Theoretical Physics,\\
University of California, Santa Barbara, CA 93106, USA
\\ \vspace{6pt}
\end{center}
\vfill
\centerline{\bf Abstract}
The dynamics of a (super)particle near the horizon of an extreme
Reissner-Nordstr\"om black hole is shown to be governed by
an action that reduces to a (super)conformal mechanics model
in the limit of large black hole mass.
\vfill
 \hrule width 5.cm
\vskip 2.mm
{\small
\noindent $^\dagger$ Onderzoeksdirecteur FWO, Belgium \\
\noindent $^\star$ On leave from DAMTP, Univ. of Cambridge, UK}
\end{titlepage}
\section{Introduction}
A new class of interacting $(p+1)$-dimensional conformal field
theories has recently been discovered as the world-volume field
theories on `test' $p$ branes in the $d$-dimensional near-horizon
background of other branes \cite{conf}. The
key point is the fact that the near-horizon geometry is of the form
$adS_{p+2}\times S^{d-p-2}$, with the $adS$ isometries being realized on
the test brane as conformal symmetry. Perhaps the simplest
realization of this idea is provided by a charged point particle near
the horizon of a $d=4$ extreme Reissner-Nordstr\"om (RN) black hole.
Here we use this example to elucidate some surprising connections
between black holes and conformal invariance.

As an illustration of the issues, consider the conformal mechanics
model of
\cite{DFF} (see also \cite{Case}) for the conjugate pair $(p,x)$. The
Hamiltonian is
\begin{equation}\label{dffham}
H= {p^2\over 2m} + {g\over 2x^2}\,.
\end{equation}
This was shown in \cite{DFF} to a have a continuous spectrum of energy
eigenstates with energy eigenvalue $E>0$, but there is no ground state
at $E=0$. In the black hole interpretation of the model, the
classical analog
of an eigenstate of $H$ is an orbit of a timelike Killing vector
field $k$,
equal to $\partial/\partial t$ in the region outside the horizon,
and the energy
is then the  value of $k^2$. The absence of a ground state of $H$ at
$E=0$ can
now be interpreted as due to the fact that the orbit of $k$ with
$k^2=0$ is a
null geodesic generator of the event horizon, which is not covered
by the static
coordinates adapted to $\partial_t$. The procedure used in \cite{DFF} to
cure this problem was to choose a different combination of conserved
charges as
the Hamiltonian. This corresponds to a different choice of time, one
for which
the worldlines of static particles pass through the black hole
horizon instead
of remaining in the exterior spacetime.

Thus, the study of conformal quantum mechanics has potential
applications to the
quantum mechanics of black holes. Here we shall limit ourselves to
an exposition
of the classical aspects of this connection, and its supersymmetric
extension.
We start from the extreme RN metric in isotropic coordinates
\begin{equation}
ds^2 = -\left(1+ {M\over \rho}\right)^{-2}dt^2 + \left(1+ {M\over
\rho}\right)^2\big[d\rho^2 + \rho^2 d\Omega^2\big]\,,
\end{equation}
where $d\Omega^2=d\theta^2 + \sin^2\theta \, d\varphi^2$ is the
$SO(3)$-invariant
metric on $S^2$, and $M$ is the black
hole mass, in units for which $G=1$. The near-horizon geometry is
therefore
\cite{CG}
\begin{equation}
ds^2 = -\left({\rho\over M}\right)^2 dt^2 + \left({M\over \rho}\right)^2
d\rho^2
+M^2d\Omega^2\,,
\end{equation}
which is the Bertotti-Robinson (BR) metric \cite{LC}. It can be
characterized as the $SO(1,2)\times SO(3)$ invariant
conformally-flat metric on
$adS_2\times S^2$. The parameter $M$ may now be interpreted as the
$S^2$ radius
(which is also proportional to the radius of curvature of the
$adS_2$ factor). A
test particle in this near-horizon geometry provides a model of conformal
mechanics in which the $SO(1,2)$ isometry of the background spacetime is
realized as a one-dimensional conformal symmetry. If the particle's
mass $m$
equals the absolute value of its charge $q$ then this is just the
$p=0$ case of
the construction of \cite{conf}. However, there is nothing to
prevent us from
considering $m\ne |q|$ and we shall begin by considering this more general
case. We shall see that this leads to a new `relativistic' model of
conformal
mechanics. In the `non-relativistic' limit, which can be viewed as a
limit of large black hole mass, one recovers the Hamiltonian
(\ref{dffham}).

Various supersymmetric generalizations of conformal mechanics have
been studied
by Akulov and Pashnev and by Fubini and Rabinovici \cite{AP}. A
`relativistic' generalization of one such model can be obtained from
the radial
dynamics of a superparticle in the near-horizon geometry of an extreme RN
solution of $d=4$ $N=2$ supergravity. An important feature of the
supersymmetric case is that the superparticle has a fermionic
gauge invariance, `$\kappa$-symmetry', when $m=|q|$. Since this
reduces the
total number of fermions by half it leads to a considerable
simplification of
the Hamiltonian governing radial motion. To take advantage of
this simplification we shall consider here only the $m=|q|$ superparticle.

\section{Conformal mechanics and black holes}

In horospherical coordinates $(t,\phi=\rho/M)$ for $adS_2$, the
4-metric and Maxwell 1-form of the BR solution of Maxwell-Einstein
theory are
\begin{eqnarray}
ds^2 &=& -\phi^2 dt^2 + {M^2\over \phi^2}\, d\phi^2 + M^2 d\Omega^2\,,
\nonumber\\
A &=& \phi dt\,.
\end{eqnarray}
The metric is singular at $\phi=0$, but this is just a coordinate
singularity and $\phi=0$ is actually a non-singular degenerate
Killing horizon of
the timelike Killing vector field $\partial/\partial t$. We now define a
new radial coordinate $r$ by
\begin{equation}
\phi = (2M/r)^2\, .
\end{equation}
The BR metric is then
\begin{equation}
ds^2 = -(2M/r)^4 dt^2 + (2M/r)^2 dr^2 + M^2d\Omega^2 \, .
\end{equation}
Note that the Killing horizon in these coordinates is now at $r=\infty$.

The (static-gauge) Hamiltonian of a particle of mass $m$ and charge
$q$ in this
background is $H=-p_0$ where $p_0$ solves the mass-shell constraint
$(p-qA)^2 + m^2 =0$. This yields
\begin{equation}
H = (2M/r)^2 \big[ \sqrt{m^2 + (r^2p_r^2 + 4L^2)/4M^2} -q\big]\,,
\end{equation}
where $L^2=p_\theta^2 + \sin^{-2}\theta\, p_\varphi^2$, which
becomes minus the
Laplacian upon quantization (with eigenvalues $\ell(\ell+1)$ for integer
$\ell$). We can rewrite this Hamiltonian as
\begin{equation}
H= {p_r^2\over 2f} + {mg\over 2r^2 f}\,,
\end{equation}
where
\begin{equation}
f = {1\over2} \big[ \sqrt{m^2 + (r^2p_r^2 + 4L^2)/4M^2} + q\big]\, ,
\end{equation}
and
\begin{equation}\label{gee}
g= 4M^2(m^2-q^2)/m + 4L^2/m\,.
\end{equation}
This Hamiltonian defines a new model of conformal mechanics. The full
set of generators of the conformal group are
\begin{equation}\label{fullgen}
H= {1\over 2f} p_r^2 + {g\over 2r^2f}\,, \qquad K = - {1\over2} fr^2\,,
\qquad D={1\over2} rp_r\,,
\end{equation}
where $K$ generates conformal boosts\footnote{Also called the generator of
`special conformal' or `proper conformal' transformations.} and $D$
generates
dilatations.
It may be verified that the Poisson brackets of these generators
close to the
algebra of $Sl(2,R)$.

To make contact with previous work on this subject, we restrict to angular
quantum number $\ell$ and consider the limit
\begin{equation}
M\rightarrow \infty\,, \qquad (m-q)\rightarrow 0\,,
\end{equation}
with $M^2(m-q)$ kept fixed. In this limit $f\rightarrow m$, so
\begin{equation}
H= {p_r^2\over 2m} + {g\over 2r^2}\,,
\end{equation}
with
\begin{equation}
g= 8M^2(m-q) + 4\ell(\ell+1)/m\,.
\end{equation}
This is the conformal mechanics of \cite{Case,DFF}. For obvious reasons we
shall refer to this as `non-relativistic' conformal mechanics; the \lq
non-relativistic' limit can be thought of as a limit of large black
hole mass.
When $\ell=0$ an \lq ultra-extreme' $m<q$ particle corresponds to
negative $g$
and the particle falls to $r=0$, i.e. it is repelled to
$\phi=\infty$. On the
other hand, a \lq sub-extreme' $m>q$ particle is pushed to
$r=\infty$, which
corresponds to it falling through the black hole horizon at
$\phi=0$. The force
vanishes (again when $\ell=0$) for an \lq extreme' $m=q$  particle,
this being a
reflection of the exact cancellation of gravitational attraction and
electrostatic repulsion in this case. A static
extreme particle of zero angular momentum follows an orbit of
$\partial/\partial t$, and remains outside the black hole horizon.

\section{Superconformal mechanics}

The `non-relativistic' conformal mechanics described above was extended in
\cite{AP} to an $SU(1,1|1)\cong OSp(2|2)$ invariant superconformal
mechanics.
This can be truncated, for $g=0$, to an $OSp(1|2)$ invariant
superconformal
mechanics, which we shall recover here as the `non-relativistic'
a limit of a `relativistic' superconformal mechanics describing the radial
motion of  a {\sl superparticle} with zero orbital angular momentum in the
near-horizon geometry of the extreme RN solution of $d=4$ $N=2$
supergravity. It
follows from the formula (\ref{gee}) that $g=0$ for this model,
since we assume
both $m=|q|$ and $\ell=0$. As will be shown elsewhere \cite{AIPT},
the equation
of motion of the $SU(1,1|1)$-invariant superconformal mechanics with
$g\ne0$ is
the `non-relativistic' limit of the radial equation of a
superparticle with
non-zero angular momentum, but here we limit ourselves to the
simpler case of
$OSp(1|2)$ and zero angular momentum.

To define the superparticle action as an integral over the image $w$
of the
worldline in superspace, we introduce (i) the superspace frame 1-forms
$E^A=(E^a,E^{\alpha i})$ (where $\alpha=1,2$ is an $Sl(2,C)$ index
and $i=1,2$ is
an index of the $SU(2)_R$ R-symmetry group) and (ii) the superspace
Maxwell
1-form $A$. The
action may then be written as
\begin{equation}\label{spart}
S= -\int_w [m\sqrt{-g} - qA]\,,
\end{equation}
where
\begin{equation}
g = E^a\otimes_s E^b\eta_{ab}\,.
\end{equation}
This action is obviously invariant (up to surface terms) under
infinitesimal isometries of the background that leave invariant the
Maxwell
field strength 2-form $F=dA$, i.e. under transformations generated
by vector
superfields $\xi$ for which
\begin{equation}
{\cal L}_\xi g =0\ , \qquad {\cal L}_\xi F=0\, .
\end{equation}
The algebra of (anti)commutators of the vector superfields $\xi$ is, by
definition, the algebra of the `isometry group of the background'.
In this case the isometry superalgebra is that of the supergroup
$SU(1,1|2)$ with bosonic subgroup $SU(1,1)\times SU(2)$.
The $SU(1,1)\times SU(2)$ subgroup is the isometry group of
$adS_2\times S^2$. This supergroup has 8 real (4 complex)
supercharges as expected from the fact that the BR solution preserves
all supersymmetries of $d=4$ $N=2$ supergravity.
The anticommutator of these odd generators is  (in $SO(1,2)\times SO(3)$
notation)
\begin{equation}
\{Q_\alpha{}^i,\bar Q_j{}^\beta\} =
-\ft14 \delta_j{}^i (\hat \gamma^{\hat m\hat
n})_\alpha{}^\beta \hat M_{\hat m\hat n} - \ft14 \delta_\alpha{}^\beta
(\gamma'{}^{m'n'})_j{}^i M'_{m'n'}\,.
\end{equation}
The $\hat \gamma_{\hat m}$ generate the $SO(1,2)$ Clifford algebra and are
chosen to be $\hat \gamma_0 = i\sigma_3,\, \hat \gamma_1 = \sigma_1$ and
$\hat \gamma_2 = i\sigma_2$, where $\sigma_i$ are the Pauli-matrices. The
$\gamma'_{m'}$ are the Pauli-matrices generating the $SO(3)$ Clifford
algebra.  $\bar Q_i{}^\alpha$ is the Dirac conjugate of
$Q_\alpha{}^i$ in $(1,2)$ dimensions, i.e. $\bar Q_i{}^\alpha = i
[(Q^i)^\dagger \hat \gamma_2
\hat \gamma_0]^\alpha$.  The conformal $SU(1,1)$ charges $(H,K,D)$
are packaged
in $\hat M_{\hat m\hat n}$ as
\begin{equation}
H=-P_0 = - 2(M_{02} + M_{01});\qquad K = 2(M_{02} - M_{01});\qquad
D=2M_{21}\,
\end{equation}
and $M'_{m'n'}$ are $SO(3)$-generators.

We now define
\begin{equation}
{\cal Q}_\alpha = Q_\alpha{}^1 + \varepsilon_{\alpha\beta} \bar
Q_1{}^\beta
+ Q_\alpha{}^2 + \varepsilon_{\alpha\beta} \bar Q_2{}^\beta\,,
\end{equation}
and it follows that
\begin{equation}
{\cal Q}_\alpha = \ba{c} S\\iQ\ea\,,
\end{equation}
where $Q$ and $S$ are real. The anticommutator of these odd generators
is
\begin{equation}
{}\{ {\cal Q}_\alpha, {\cal Q}_\beta\} = - M_{\alpha\beta}\,,
\end{equation}
where
\begin{equation}
M_{\alpha\beta} = \ba{cc} iK & D\\D & iH\ea\,.
\end{equation}
Thus the charges $(H,K,D,Q,S)$ generate a sub-supergroup which is actually
$OSp(1|2;R)$ (the non-vanishing (anti)commutation relations are given in
(\ref{OSP12}) below). This is the sub-supergroup relevant to the
truncated
system in which we consider a superparticle moving radially. This
system is
equivalent to a $d=2$ superparticle on a superspace with $adS_2$
`body' and
isometry supergroup $OSp(1|2;R)$, the $Sp(2;R)\cong SU(1,1)$
subgroup being the
isometry group of $adS_2$. This simplified model still captures the
essential
feature of the black hole, i.e. the existence of an event horizon.

One has only to gauge fix the reparametrization invariance of the
action for a
superparticle in this $adS_2$ superspace to find a model of superconformal
mechanics, but unless $m=q$, both the standard supersymmetry and the
conformal
supersymmetry will be non-linearly realized, i.e. there will be no state
annihilated by either $Q$ or $S$. This is hardly surprising since there is
clearly no classical solution of zero energy when $g\ne 0$ whereas
there is when
$g=0$. This distinction is reflected in the \lq $\kappa$-symmetry'
of the $m=|q|$
action which, for reasons explained in detail elsewhere, ensures
that half of
the supersymmetries are linearly realized. In the present context,
it means that
$Q$ is linearly realized in that the ground state is annihilated by
$Q$, while
$S$ is non-linearly realized. This is the case that we are going to
study in
detail in this paper.

We proceed by first passing to the Hamiltonian form of the above
superparticle
action, which is a functional of the $(2|2)$ superspace coordinate
variables
$Z^M$ and their conjugate momenta $p_M$. The Lagrangian in this form is
\begin{equation}
L = \dot Z^M p_M -{1\over2} v(\tilde p^2 + m^2) +
\zeta^\alpha E_\alpha{}^M (p_M -qA_M)\,,
\end{equation}
where $v$ is a Lagrange multiplier for the mass-shell constraint,
$\zeta$ is
a two-component real spinor Lagrange multiplier for the fermionic
constraints,
and
\begin{equation}
\tilde p_a = E_a{}^M(p_M - qA_M)\, .
\end{equation}
The fermionic constraints are purely second class if $m\ne q$, but half
first-class and half second-class when $m=q$. Now, $E_a{}^\mu$
vanishes in flat
superspace. It must therefore continue to vanish in any
superconformally-flat
superspace since the supervielbeins are obtained in such cases from
that of flat
superspace by a super-Weyl transformation with scalar superfield parameter
\cite{howe}. The BR background is superconformally flat, so we have
\begin{equation}
\tilde p ^2= g^{mn}(p_m - qA_m)(p_n - qA_n) \, ,\qquad
(g^{mn} \equiv \eta^{ab}E_a{}^m E_b{}^n).
\end{equation}
The mass-shell constraint for the superparticle is therefore  
formally identical
to that of the bosonic particle. The only difference resides in the  
fact that
the inverse metric $g^{mn}$ and the Maxwell 1-form $A_m$ are {\sl
superfields}. Their leading components are just the inverse metric  
and Maxwell
1-form of the bosonic action, but they will also contain terms  
proportional to
fermions.

Now, all fermion terms in the expansion of $g^{mn}$ and $A_m$ must  
be even in
fermions. In the special case that the superspace is $(2|2)$  
dimensional with
$adS_2$ body the expansion in fermions must terminate at the  
quadratic order
because there are only two fermionic variables. If we further  
specialize to the
$m=|q|$ case then only one combination of these two can actually appear
(this is implied by $\kappa$-invariance). Thus, all fermion  
bilinears vanish
identically in this case and the mass-shell constraint, and hence the
Hamiltonian, is {\sl identical} to that of the bosonic particle. The
same is true of all the $Sl(2;R)$ generators. The remaining generators of
$OSp(1|2;R)$ are the supersymmetry charge $Q$ and the generator of
superconformal boosts (alias `special' supersymmetry) $S$. These could be
deduced from the charges associated with the fermionic Killing vector
superfields of the background superspace, but it is easy to guess
them as they
are necessarily linear in the one physical fermion, which we may
call $\psi$.
The final result is as follows. The $Sp(2;R)\cong Sl(2;R)$
generators of this
($m=q$, $d=2$) model are
\begin{equation}
H= {1\over 2f} p_r^2\,,  \qquad K = -{1\over2} f r^2\,,
\qquad D= {1\over2} rp_r\,,
\end{equation}
where
\begin{equation}
f = {1\over2}m \big[ \sqrt{1 + (rp_r/2mM)^2} + 1\big] \ ,
\end{equation}
and the fermionic generators are
\begin{equation}
Q = {p_r\over\sqrt{2f}}\,\psi\,, \qquad S= \sqrt{f/ 2}\, r\psi\,,
\end{equation}
where $\psi$ is an anticommuting worldline `field'. Given the  
Poisson bracket
relations
\begin{equation}
\{r,p_r\}=1\,, \qquad \{\psi,\psi\}= i\,,
\end{equation}
one may verify that these generators define the Lie superalgebra of
$OSp(1|2;R)$. Specifically, the non-zero PB relations are
\begin{eqnarray}
&&\{D,H\} = H\,, \hspace{1.5cm}  \{D,K\} =
-K\,,\hspace{1.5cm}\{H,K\} = 2D\,,
\nonumber\\
&&\hspace{.05cm}\{D,Q\} = \ft12 Q\,,\hspace{1.42cm}
\{D,S\}= - \ft12 S\,,\nonumber \\
&&\hspace{.1cm}\{H,S\}=-Q \,,  \hspace{1.27cm}  \{K,Q\} = -S\,,
\nonumber \\
&&\hspace{.08cm}\{Q,Q\}= iH \,, \hspace{1.56cm} \{S,S\}
=-iK\,,\hspace{1.4cm}
\{Q,S\} = iD\,.\label{OSP12}
\end{eqnarray}
In the $M\rightarrow\infty$ limit we obtain an $OSp(1|2)$ invariant
superconformal mechanics model with $g=0$.

\section{Discussion}
We have shown that the dynamics of a (super)particle in the near-horizon
geometry of the extreme RN solution of $d=4$ $N=2$ supergravity is
governed by a model of (super)conformal mechanics that generalizes
previous
constructions of such models. For purely radial motion, ($L^2=0$) and when
$m=|q|$ there is a family of degenerate ground states of the
particle Hamiltonian
parametrized by $\langle r\rangle$. Because $r$ scales under dilatations,
conformal invariance is spontaneously broken for any finite or non-zero
$\langle r\rangle$, but it is unbroken when either $\langle r\rangle=0$ or
$\langle r\rangle=\infty$. As explained in a slightly different context
in \cite{conf}, the quantity $\langle r\rangle/M$ is effectively the
coupling constant, so the `end of the universe' limit $\langle r\rangle
\rightarrow 0$ (recall that this corresponds to
$\langle\phi\rangle\rightarrow
\infty$) is equivalent to the $M\rightarrow\infty$ limit in which we
 obtain a
free non-relativistic superconformal mechanics. The other limit in which
$\langle r\rangle\rightarrow \infty$ is an ultra-relativistic one in
which the
particle's orbit approaches a null geodesic generator of the Killing
horizon.
The Hamiltonian governing the particle's dynamics in this limit may
be found by taking $M\rightarrow 0$ for fixed $m$ and $q$. In the
$L^2=0$ case
this yields
\begin{equation}
H = {2Mp_r\over r} + {\cal O}(M^2)\,.
\end{equation}
By ignoring the ${\cal O}(M^2)$ terms we effectively take the limit,
and the
$Sl(2,R)$ generators reduce to
\begin{equation}
H = {2Mp_r\over r}\,, \qquad K = -{r^3p_r\over 8M}\,,
\qquad D = {1\over2}rp_r\,.
\end{equation}
The $M$-dependence may now be removed by the rescaling $r\rightarrow
\sqrt{M} r$,  $p_r\rightarrow p_r/\sqrt{M}$. The absence of any
dependence of this Hamiltonian on $m$ and $q$ means that the full
symmetry group
of this model is that of the {\sl massless} (super)particle in the same
background. For superconformally flat backgrounds, such as $adS_2$
or the BR
spacetime, the symmetry group is the same as that of a free particle
in flat
space, and is therefore an infinite rank extension of the
superconformal group
\cite{PKT}.
\medskip
\section*{Acknowledgements.}
\noindent
We thank M. Halpern, and a referee, for bringing ref. \cite{Case}
and ref. \cite{LC} to our attention   and E. Rabinovici for enlightening
discussion.
The work of R. K. and J. K. is supported by the NSF grant PHY-9219345.
The work of J. K. is also supported by the United States
Department of Defense, NDSEG Fellowship program.  This work is
supported by the European Commission TMR programme
ERBFMRX-CT96-0045.

\end{document}